\documentclass{iopconfser}
\usepackage{graphicx}
\usepackage{cite}
\begin{document}

\title{Translational Symmetry Broken Magnetization Plateau of the $S={{1}\over{2}}$ Anisotropic Spin Ladder with Ferromagnetic Rung Interaction}

\author{T\^oru Sakai$^{1,2}$, Koki Doi$^{1}$, Kiyomi Okamoto$^1$, Kouichi Okunishi$^3$, Masaru Hashimoto$^1$, Tomoki Houda$^1$, Rito Furuchi$^1$ and Hiroki Nakano$^{1}$}

\affil{$^1$School of Science, University of Hyogo, Hyogo 678-1297, Japan}
\affil{$^2$National Institutes for Quantum Science and Technology, SPring-8, Hyogo 679-5148, Japan}
\affil{$^3$Department of Physics, Niigata University, Niigata 950-2181, Japan}

\email{sakai@spring8.or.jp}

\begin{abstract}
The magnetization process of the $S=1/2$ anisotropic spin ladder with the ferromagnetic rung interaction is investigated using the numerical diagonalization of finite-size clusters. 
It is found that the translational symmetry broken magnetization plateau would appear at half the saturation magnetization, when the competing anisotropies are sufficiently large. 
The phase diagram with respect to the anisotropies and several magnetization curves are presented. 
\end{abstract}

\section{Introduction}
The magnetization plateau has attracted a lot of interest in the research field of magnetism. 
It would be observed as the plateau of the magnetization curve, when the external magnetic field induces the spin gap. 
Based on the Lieb-Schulz-Mattis theorem, Oshikawa, Yamanaka and Affleck\cite{oshikawa} derived the necessary condition 
of the appearance of the magnetization plateau in the one-dimensional spin systems as the form $Q(S-\tilde m)={\rm integer}$, 
where $S$ and $\tilde m$ are the spin and magnetization per unit cell, and $Q$ is the periodicity of the ground state. 
For example, $Q$ should be two for the Neel ordered state. 
The magnetization plateau for $Q=1$ was theoretically predicted to appear in the $S=3/2$ and 2 antiferromagnetic chains 
with the single-ion anisotropy, using the numerical diagonalization of
finite-size systems\cite{sakai1,kitazawa-okamoto,sakai2}.
In the distorted diamond chains $Q=1$ plateau was found both theoretically and experimentally
\cite{honecker1,okamoto1,kikuchi,gu-su,honecker2,ananikian,morita,ueno,filho}. 
Other examples are listed in Ref.\cite{sakai3}.
The plateau for $Q=2$ was also indicated to occur in many geometrically frustrated systems
\cite{totsuka,okunishi1,okunishi2,metavitsiadis,gong,mahdavifar,jiang,nakano,
sugimoto1,sugimoto2,sasaki,okamoto2,okamoto3,michaud,kohshiro}. 
Recently, using the numerical diagonalization and the density matrix renormalization group analyses, the plateau for $Q=2$ was predicted to appear 
in the $S=1$ antiferromagnetic chain with the competing anisotropies without the geometric frustrations,
namely the easy-axis coupling anisotropy and the easy-plane single-ion one\cite{sakai3}. 
This mechanism of the plateau based on the competing anisotropies is also expected to work in the $S=1/2$ spin ladder with the ferromagnetic rung interaction, 
as well as the ferromagnetic and antiferromagnetic bond-alternating chain\cite{sakai4}. 
In this paper, we investigate the possibility of the 1/2 magnetization plateau in the $S=1/2$ spin ladder 
with the easy-axis and easy-plane anisotropies introduced to the antiferromagnetic leg and ferromagnetic rung interactions, respectively. 
The numerical diagonalization of finite-size clusters
and the level spectroscopy analysis\cite{okamoto-nomura,nomura-okamoto} 
will indicate that the 1/2 magnetization plateau appears for sufficiently large anisotropies. 

\section{Model and calculations}

The magnetization process of the $S=1/2$ spin ladder with the coupling anisotropies $\lambda$ and $\gamma$ introduced to the ferromagnetic rung 
and the antiferromagnetic leg interactions, respectively, is described by the Hamiltonian
\begin{eqnarray}
\label{ham}
{\cal H}&=&{\cal H}_0+{\cal H}_Z, \\
\nonumber
{\cal H}_0& =& J_{\rm r} \sum_{j=1}^L\left[S_{1,j}^xS_{2,j}^x + S_{1,j}^yS_{2,j}^y  + \lambda S_{1,j}^zS_{2,j}^z  \right] 
 +J_1\sum_{i=1}^2 \sum_{j=1}^L\left[\gamma(S_{i,j}^xS_{i,j+1}^x + S_{i,j}^yS_{i,j+1}^y)  +  S_{i,j}^zS_{i,j+1}^z  \right] \\
\nonumber
{\cal H}_Z& =&-H\sum_{i=1}^2 \sum_{j=1}^L S_{i,j}^z 
\end{eqnarray}
where $S_{i,j}^\mu$ is the $\mu$ component of the $S=1/2$ spin operator at the $j$th site of the $i$th leg,
and $H$ is the external magnetic field. 
The ferromagnetic rung interaction $J_{\rm r}$ is fixed to $-1$. 
We consider the case of the competing anisotropies; the easy-plane $\lambda(<1)$ and the easy-axis $\gamma(<1)$. 
Using the Lanczos method, we calculate the lowest energy $E(L,M)$ in the subspace where $\sum _j S_{j}^z=M$ 
for each $L$, under the periodic boundary condition. 
The reduced magnetization $m$ is defined by $m=M/M_{\rm s}$, 
where $M_{\rm s}$ denotes the saturation of the magnetization, 
namely $M_{\rm s}=L$. 

\section{1/2 magnetization plateau}

\begin{figure}[ht]
\bigskip
\centerline{\includegraphics[width=0.45\linewidth,angle=0]{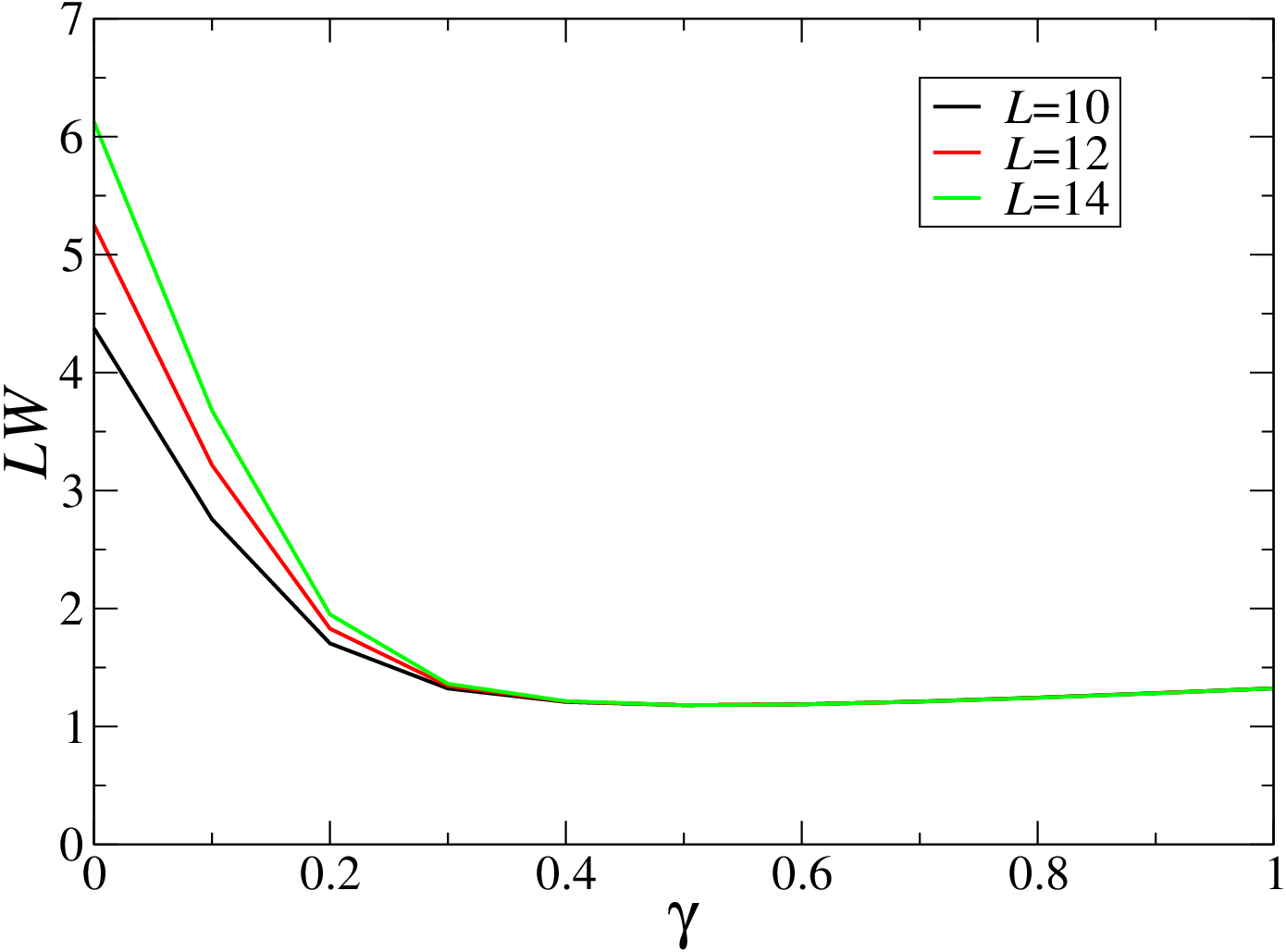}}%
\caption{\label{width}
Scaled plateau width $LW$ plotted versus $\gamma$ at $m=1/2$ for $L=10,~12$ and $14$, 
in the case of $J_1=0.5$ and $\lambda =0.5$. For sufficiently small $\gamma$, $LW$ increases 
with $L$. 
It suggests that the 1/2 magnetization plateau appears there. 
}
\end{figure}

\begin{figure}[ht]
\centerline{\includegraphics[width=0.45\linewidth,angle=0]{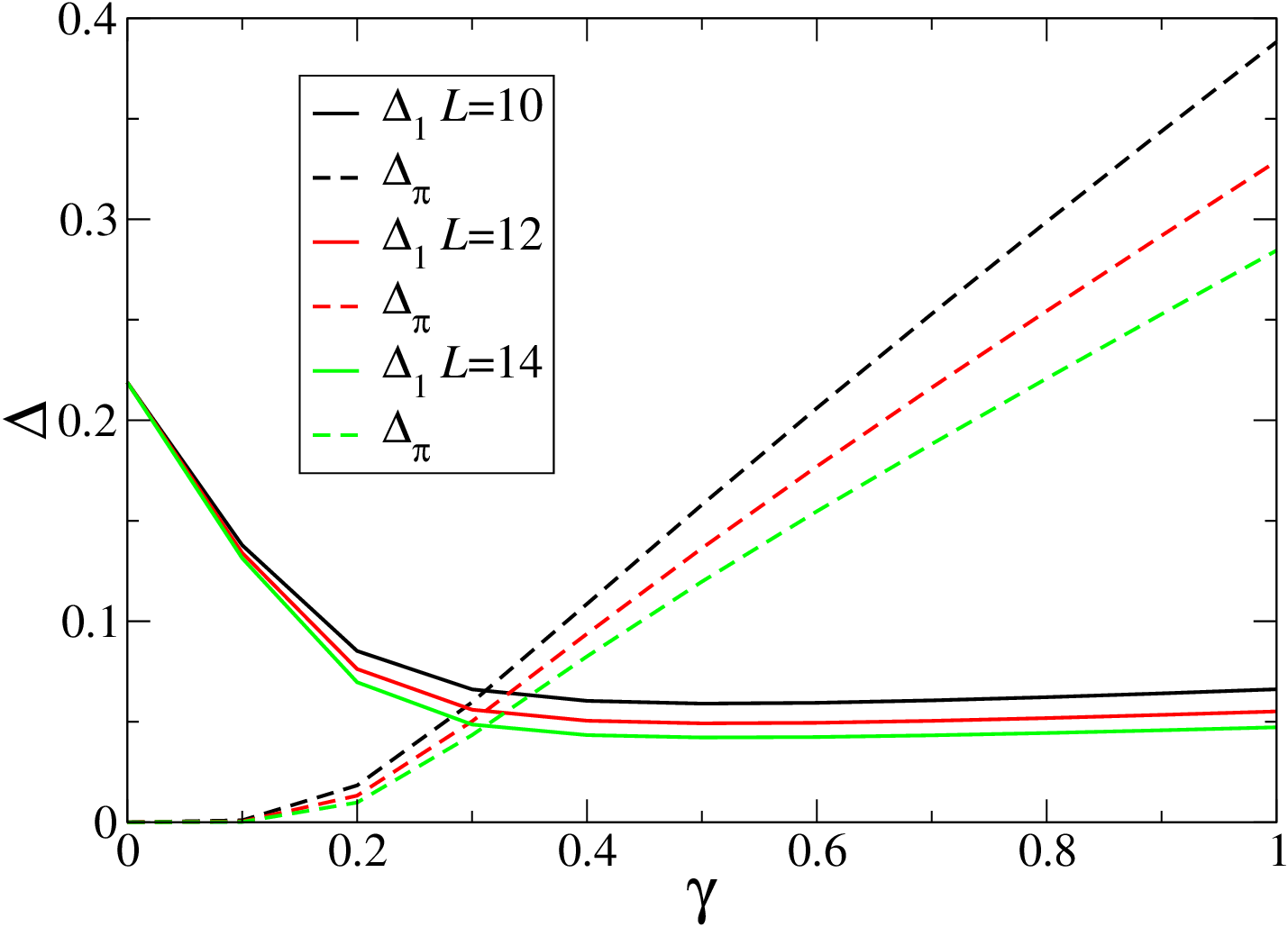}}%
\caption{\label{gap}
Excitation gaps $\Delta _1$ and $\Delta _{\pi}$ plotted versus $\gamma$ at $m=1/2$ for $L=10,~12$ and $14$, 
in the case of $J_1=0.5$ and $\lambda =0.5$. The cross point of $\Delta _1$ and $\Delta _{\pi}$ 
in the limit $L\rightarrow \infty$ gives the boundary of the plateau phase. 
}
\end{figure}

\begin{figure}[ht]
\bigskip
\centerline{\includegraphics[width=0.50\linewidth,angle=0]{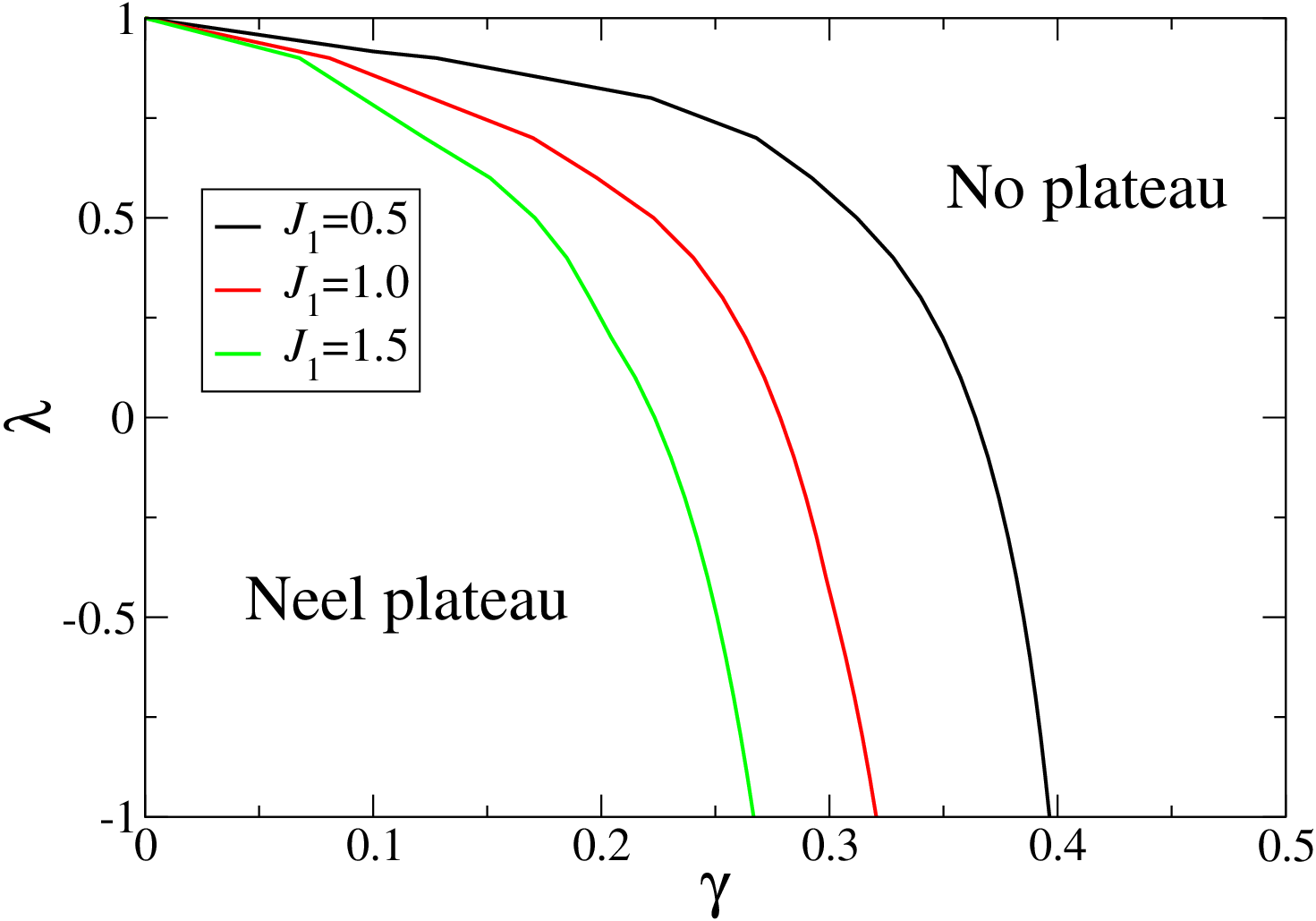}}%
\caption{
\label{phase}
Phase diagram at $m=1/2$ with respect to the two anisotropies $\gamma$ and $\lambda$. }
\end{figure}

We focus on the magnetization plateau at $m=1/2$. 
If it appears, $Q=2$ should be satisfied. 
In the case of $J_1=0.5$ and $\lambda=0.5$, 
we calculated the plateau width for finite-size systems defined by $W=E(L,M+1)+E(L,M-1)-2E(L,M)$, 
where $M=L/2$. 
The scaled plateau width $LW$ is plotted versus $\gamma$ for $L=10,~12$ and $14$ in Fig. \ref{width}. 
It indicates that the plateau width would be finite even in the infinite $L$ limit for sufficiently small $\gamma$. 
We use the level spectroscopy analysis\cite{okamoto-nomura,nomura-okamoto} to estimate the boundary of the plateau phase at $m=1/2$. 
According to this method, the cross point of the two excitation gaps $\Delta _1$ and $\Delta _{\pi}$ 
gives a good estimation of the phase boundary. 
Here $\Delta _1$ and $\Delta _{\pi}$ are defined by 
$\Delta _1 \equiv [E(L,M+1)+E(L,M-1)-2E(L,M)]/2$ and $\Delta _{\pi} \equiv E_{k=\pi}(L,M)-E(L,M)$ where 
$E_{k=\pi}(L,M)$ is the lowest excitation energy with the wave vector $k=\pi$. 
These gaps for $J_1=0.5$ and $\lambda =0.5$ are plotted versus $\gamma$ for $L$=10, 12 and 14 
in Fig. \ref{gap}. 
The $L$ dependence of the cross points of $\Delta _1$ and $\Delta _{\pi}$ is quite small. 
Assuming the finite-size correction proportional to $1/L^2$, 
we estimate the boundary of the plateau phase in the infinite $L$ limit. 
Applying this method, we obtain the phase diagrams with respect to the two anisotropies $\gamma$ and $\lambda$ 
for $J_1=0.5,~1.0$ and $1.5$, which are shown in Fig. \ref{phase}. 
The relation $\Delta _1 > \Delta _{\pi}$ indicates the 2-fold degeneracy in the ground state, 
namely $Q=2$. 
Probably each rung pair would form the Neel-like order $|\cdots 010101 \cdots \rangle$. 
Then we call this plateau Neel plateau. 
When $J_1$ increases, the plateau phase shrinks. 


\section{Magnetization curves}

\begin{figure}[ht]
\centerline{\includegraphics[width=0.45\linewidth,angle=0]{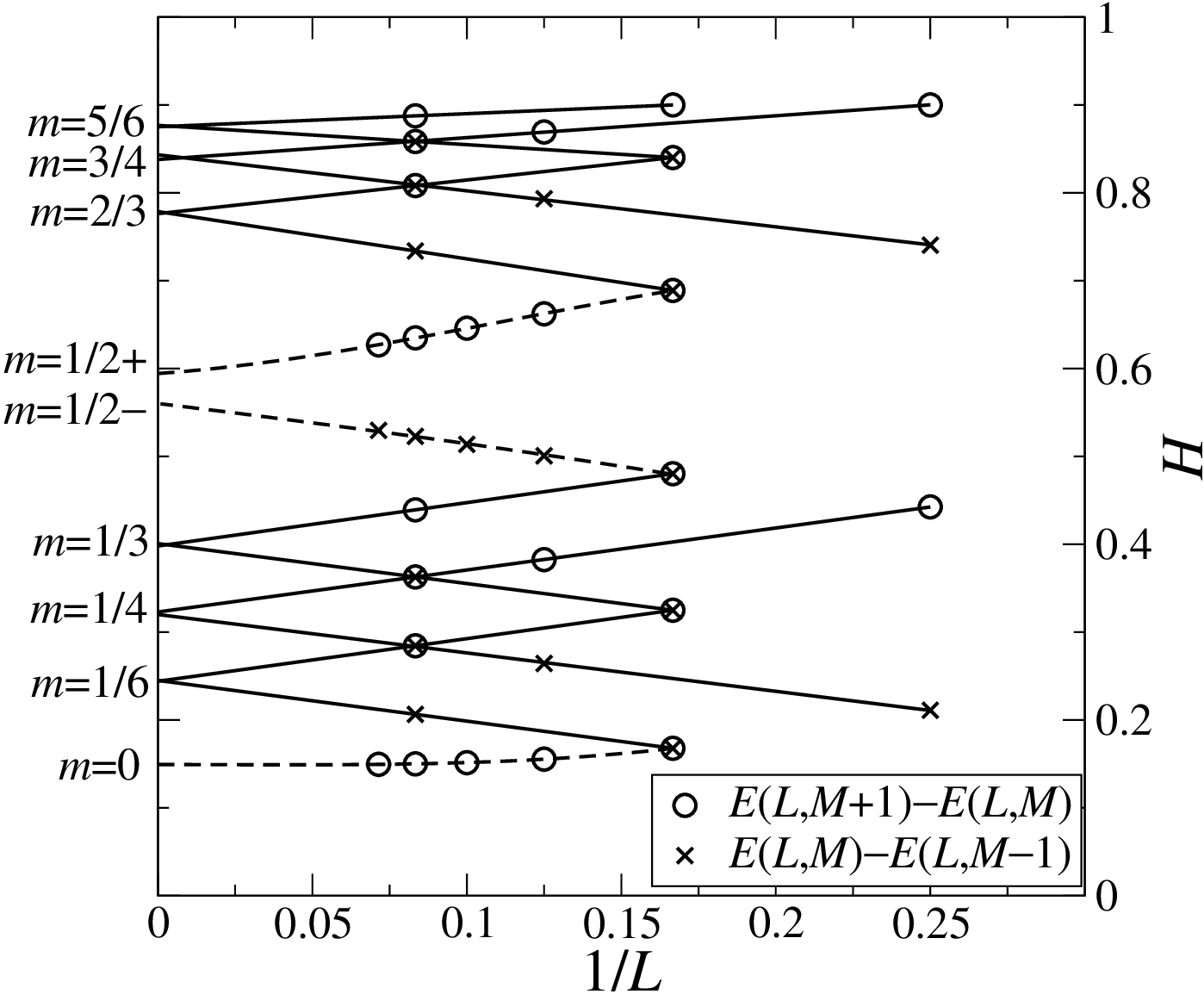}}%
\bigskip
\caption{\label{extra}
 $E(L,M)-E(L,M-1)$(crosses) and 
$E(L,M+1)-E(L,M)$(circles) plotted versus $1/L$ for $J_1=0.5, \lambda=0.5$ and 
$\gamma=0.3$. Each two quantities seem to coincide to the magnetic field $H$ for $m$ 
in the infinite $L$ limit except for $m=0$ and $1/2$. 
The extrapolated points for $m=0$, $m=1/2-$ and $m=1/2+$ correspond to the 
results of the Shanks transformation. 
}
\end{figure}

It is useful to obtain the theoretical magnetization curve for 
several typical anisotropy parameters. 
The magnetic fields $H_-(m)$ and $H_+(m)$ are defined as 
\begin{eqnarray}
E(L,M)-E(L,M-1) \rightarrow H_-(m) \quad (L\rightarrow \infty), \\
E(L,M+1)-E(L,M) \rightarrow H_+(m) \quad (L\rightarrow \infty) , 
\label{mag}
\end{eqnarray}
where the size $L$ is varied with fixed $m=M/M_s$. 
If the system is gapless at $m$, the size correction should be 
proportional to $1/L$ and $H_-(m)$ coincides to $H_+(m)$. 
It is justified by Fig. \ref{extra}, where $E(L,M)-E(L,M-1)$ and 
$E(L,M+1)-E(L,M)$ are plotted versus $1/L$ for $J_1=0.5, \lambda=0.5$ and 
$\gamma=0.3$. 
It suggests that the system is gapless except for $m=0$ and $m=1/2$. 
Then we can estimate $H(m)=H_-(m)=H_+(m)$ using the form
\begin{eqnarray}
E(L,M+1)-E(L,M-1)  \rightarrow H(m) + O(1/L^2).
\label{field}
\end{eqnarray}
If the system has a gap at $m$, the magnetization plateau exists 
and $H_-(m) \not= H_+(m)$. 
In such a case, we use the Shanks transformation\cite{shanks}
\begin{eqnarray}
P'_{L}={{P_{L-2}P_{L+2}-P_L^2}\over{P_{L-2}+P_{L+2}-2P_L}}.
\label{shanks}
\end{eqnarray}
 to estimate  $H_-(m)$ and $H_+(m)$, independently. 
 The result of the Shanks transformation applied for the sequence 
  $E(L,M)-E(M-1)$ twice to estimate $H_-(1/2)$ for $j_1=0.5$, $\lambda=0.5$ and $\gamma=0.3$, 
  as shown in Table 1.

\begin{table}[ht]
\bigskip
   \caption{Result of the Shanks transformation applied for the sequence 
   $E(L,M)-E(M-1)$ twice to estimate $H_-(1/2)$ for $j_1=0.5$, $\lambda=0.5$ and $\gamma=0.3$. 
   The estimated result is $0.41 \pm 0.01$. }
   \bigskip
   \begin{tabular}{|c|c|c|c|}
      \hline
      $L$& $P_L$ & $P_L'$ &$P_L''$ \\ \hline
      ~6~& ~0.3858905~ && \\ \hline
      ~8~& 0.3947657& ~0.4066779~ & \\ \hline
      ~10~&0.3998516 &0.4087753 & ~0.4126742~ \\ \hline
     ~12~~& 0.4030912  & 0.4101391  & \\ \hline
     ~14~~&0.4053106  && \\ \hline
   \end{tabular}
   \label{shanks1}
\end{table}

\begin{figure}[ht]
\bigskip
\centerline{\includegraphics[width=0.45\linewidth,angle=0]{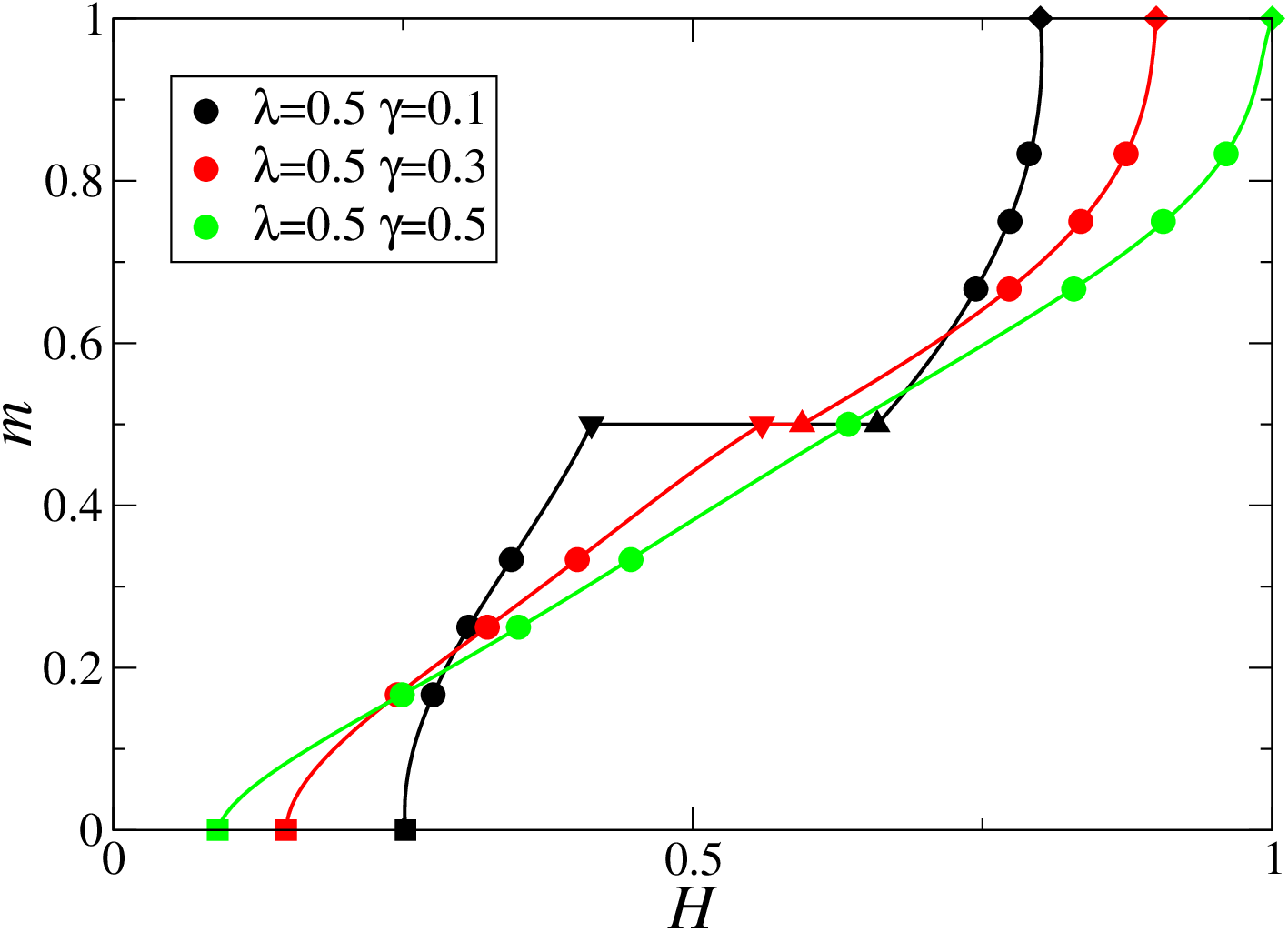}}%
\caption{\label{curve}
Magnetization curves for $J_1=0.5$ and $\lambda=0.5$. 
Black, red and green symbols are for $\gamma=$0.1, 0.3 and 0.5, respectively. 
The 1/2 magnetization plateau appears for $\gamma=$0.1 and 0.3. 
The gapless points (circles) are estimated by (\ref{field}), the gapped points 
(squares and triangles) are estimated by the Shanks transformation, 
and the saturation field (diamonds) are calculated numerically 
(They are almost independent of $L$). 
Lines are guides for the eye. 
}
\end{figure}

\begin{figure}[ht]
\bigskip
\centerline{\includegraphics[width=0.45\linewidth,angle=0]{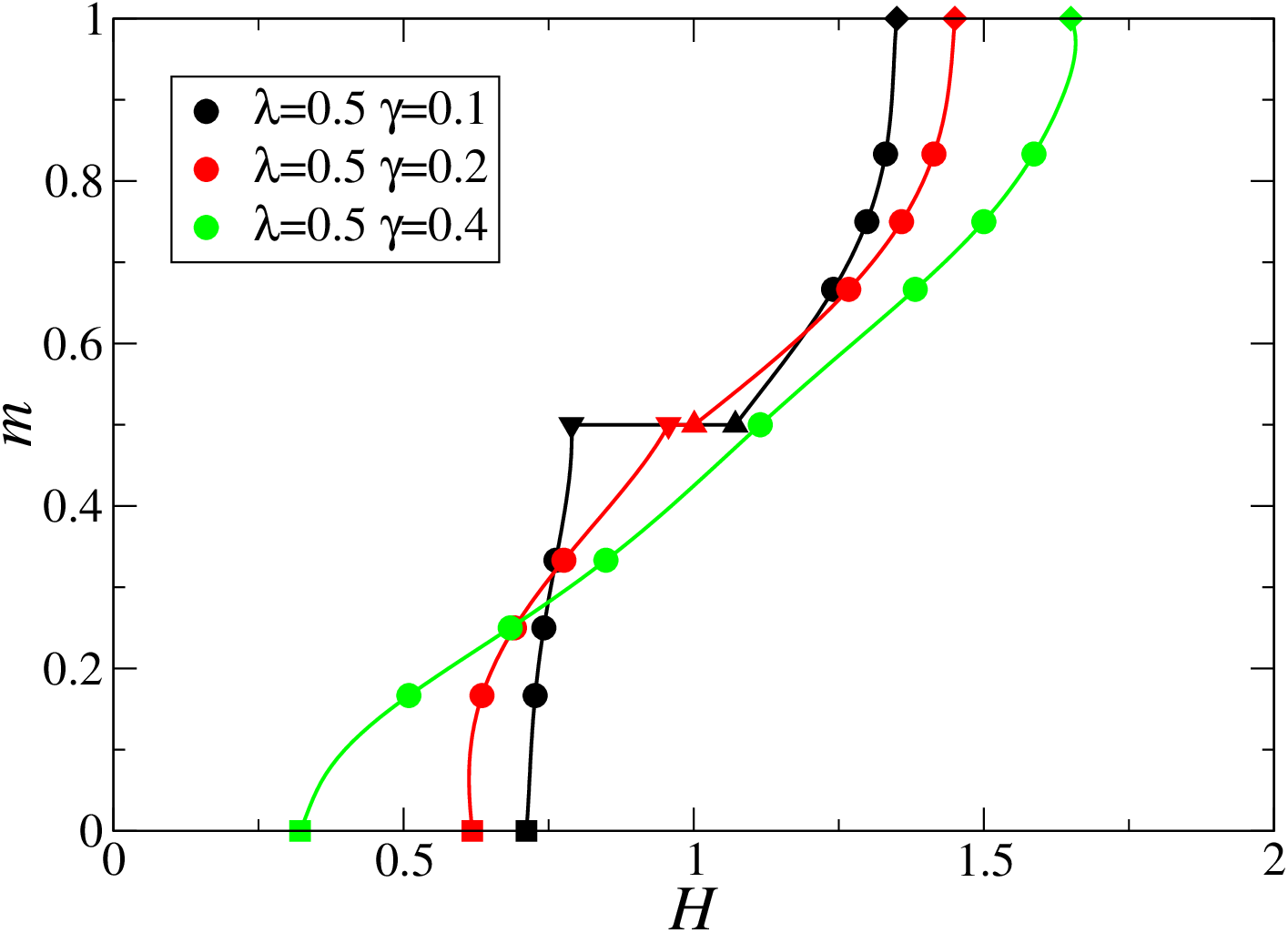}}%
\caption{\label{curve10}
Magnetization curves for $J_1=1.0$ and $\lambda=0.5$. 
Black, red and green symbols are for $\gamma=0.1,~0.2$ and $0.4$, respectively. 
The 1/2 magnetization plateau appears for $\gamma=$0.1 and $0.2$. 
The meanings of the circles, squares, triangles and diamonds are the same as those of Fig. \ref{curve}.}
\end{figure}

\begin{figure}[ht]
\bigskip
\centerline{\includegraphics[width=0.45\linewidth,angle=0]{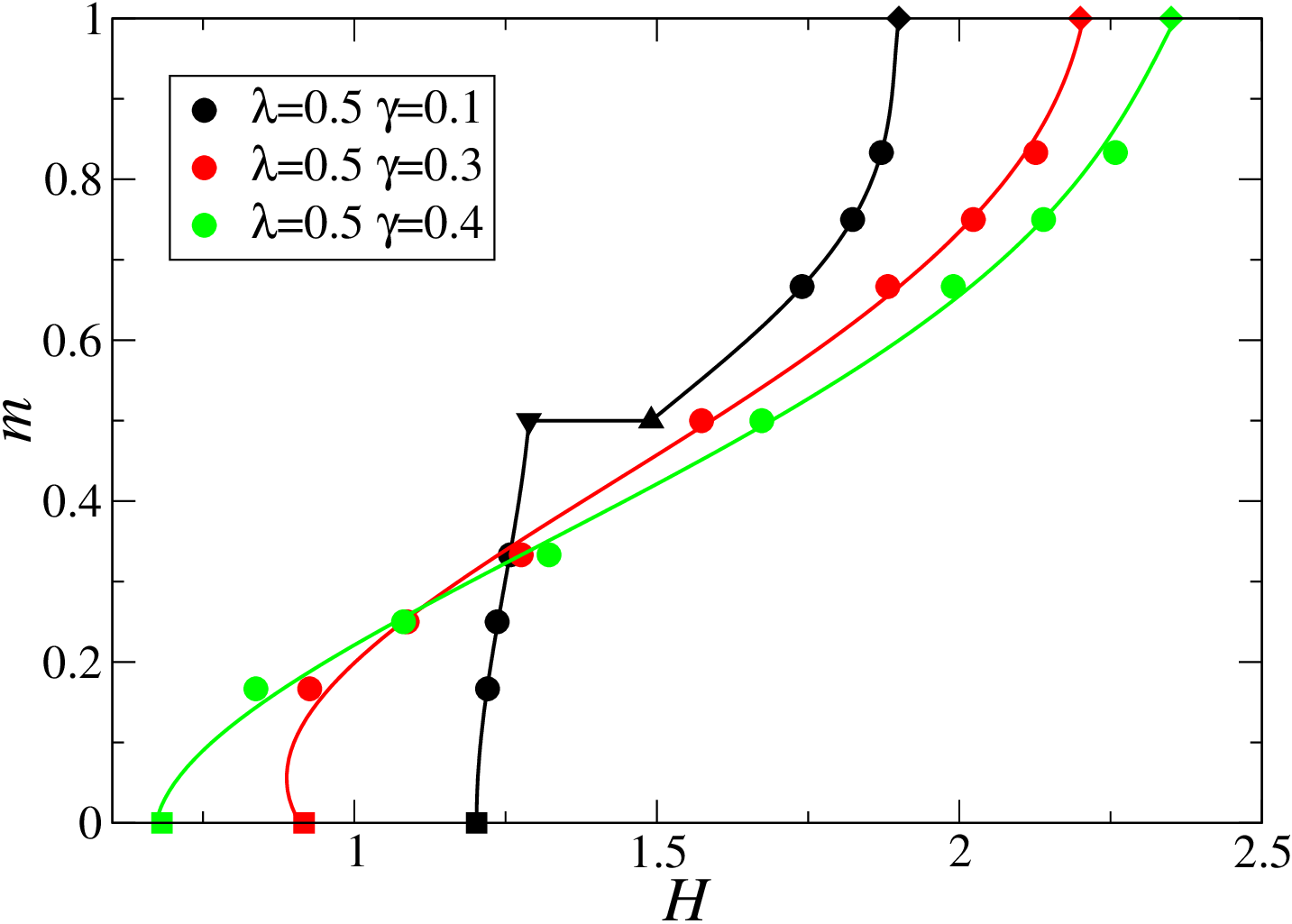}}%
\caption{\label{curve15}
Magnetization curves for $J_1=1.5$ and $\lambda=0.5$. 
Black, red and green symbols are for $\gamma=0.1,~0.3$ and $0.4$, respectively. 
The 1/2 magnetization plateau appears for $\gamma=0$.1. 
The meanings of the circles, squares, triangles and diamonds are the same as those of Fig. \ref{curve}.}
\end{figure}

 Using these methods, the magnetization curves in the infinite $L$ limit 
 for $J_1=0.5$ and $\lambda=0.5$ are obtained in Fig. \ref{curve}, 
 where circles are gapless points estimated by Eq.(\ref{field}),  squares and triangles 
 are gapped points by Shanks transformation, and diamonds are the saturation field 
 calculated numerically. 
 The lines are guides for the eye. 
 The 1/2 magnetization plateau appears for $\gamma =0.1$ and $0.3$. 
 Using the same analyses, we also obtained the magnetization curves 
 for $J_1=1.0$ and $1.5$, shown in Figs. \ref{curve10} and \ref{curve15}, 
 respectively. 
All of the shapes of the above magnetization curves are consistent with the phase diagrams of Fig.3.

\section{Summary}
The magnetization process of the $S=1/2$ spin ladder with the ferromagnetic rung and the antiferromagnetic leg interactions 
are investigated using the numerical diagonalization of finite-size clusters. 
When the easy-plane and easy-axis anisotropies are introduced to the rung and leg interactions, respectively, 
the translational symmetry broken magnetization plateau can be predicted to appear at $m=1/2$. 
The phase diagram for the two anisotropies is obtained. 
In addition several magnetization curves are presented. 
We hope such a magnetization plateau would be observed in some real materials.

\section*{Acknowledgments}
This work was partly supported by JSPS KAKENHI, 
Grant Numbers JP16K05419, JP20K03866, JP16H01080 (J-Physics), 
JP18H04330 (J-Physics), JP20H05274, JP21H05182, JP21H05191 and
23K11125, and also by JST, CREST Grant Number JPMJCR24I1.
A part of the computations was performed using
facilities of the Supercomputer Center,
Institute for Solid State Physics, University of Tokyo,
and the Computer Room, Yukawa Institute for Theoretical Physics,
Kyoto University.
We used the computational resources of the supercomputer 
Fugaku provided by the RIKEN through the HPCI System 
Research projects (Project ID: hp200173, hp210068, hp210127, 
hp210201, hp220043, hp230114, hp230532 and hp230537
).


\end{document}